\RequirePackage{fix-cm}
\documentclass{svjour3}                     % onecolumn (standard format)
\usepackage{graphicx}
\usepackage{multirow}
\usepackage{graphicx}
\usepackage{textcomp}
\usepackage{dcolumn}
\usepackage{amsmath}
\usepackage{epstopdf}
\usepackage{rotating}
\usepackage{color}

\usepackage{natbib}
\setcitestyle{aysep={}}
\usepackage{url}
\usepackage{makecell}
\usepackage{setspace}
\usepackage{etoolbox}
\usepackage{lipsum}
\usepackage{setspace}
\AtBeginEnvironment{tabular}{\singlespacing}

\begin{document}
\title{Participation and Performance on Paper- and Computer-Based Low-Stakes Assessments}
\titlerunning{Participation and Performance on PPT and CBT Low-Stakes Assessments}  
%\author{-}
%\institute{-}

\author{Jayson M. Nissen         \and
        Manher Jariwala \and 
        Eleanor W. Close \and 
        Ben {Van Dusen}
}


\institute{Jayson M. Nissen \at
              California State University Chico, 101 Holt Hall, Chico, CA, 95929, USA \\
              Tel.: 406-581-1982\\
              \email{jnissen1@csuchico.edu}           
}

\date{Received: date / Accepted: date}
\maketitle
\begin{abstract}
High-stakes assessments, such the Graduate Records Examination, have transitioned from paper to computer administration. Low-stakes Research-Based Assessments (RBAs), such as the Force Concept Inventory, have only recently begun this transition to computer administration with online services. These online services can simplify administering, scoring, and interpreting assessments, thereby reducing barriers to instructors' use of RBAs. By supporting instructors' objective assessment of the efficacy of their courses, these services can stimulate instructors to transform their courses to improve student outcomes. We investigate the extent to which RBAs administered outside of class with the online Learning About STEM Student Outcomes (LASSO) platform provide equivalent data to tests administered on paper in class, in terms of both student participation and performance. We use an experimental design to investigate the differences between these two assessment conditions with 1,310 students in 25 sections of 3 college physics courses spanning 2 semesters. Analysis conducted using Hierarchical Linear Models indicates that student performance on low-stakes RBAs is equivalent for online (out-of-class) and paper-and-pencil (in-class) administrations. The models also show differences in participation rates across assessment conditions and student grades, but that instructors can achieve participation rates with online assessments equivalent to paper assessments by offering students credit for participating and by providing multiple reminders to complete the assessment. We conclude that online out-of-class administration of RBAs can save class and instructor time while providing participation rates and performance results equivalent to in-class paper-and-pencil tests.		
\end{abstract}

\section{Introduction}
Research-based assessments (RBAs), such as the Force Concept Inventory (FCI) \citep{Hestenes1992}, the Conceptual Survey of Electricity and Magnetism (CSEM) \citep{Maloney2001}, and the Colorado Learning Attitudes about Science Survey (CLASS) \citep{Adams2006}, measure students' knowledge of concepts or attitudes that are core to a discipline. The demonstrated efficacy of RBAs in the research literature has led many instructors to use them to assess student outcomes and to develop and disseminate research-based teaching practices, particularly in the STEM disciplines \citep{DBER}. However, \citet{Madsen2016} found that instructors face several barriers to using RBAs, including choosing assessments, administering and scoring the assessments, and interpreting results. 
\par 	Educators and researchers have developed several online resources to support instructors' adoption of RBAs. A central thrust of these efforts is the development of tools to make it easy for instructors to quickly and easily collect high-quality student RBA data. For example,
\begin{enumerate}
\item \url{www.physport.org/},
\item \url{hcuboulder.qualtrics.com/jfe/form/SV_086qKlJAMx8VaMl}, and
\item \url{learningassistantalliance.org/public/lasso.php}.
\end{enumerate}
As use of online data collection systems increases, it is important to establish whether online administration of RBAs outside of class provides equivalent data to the traditional in-class, paper-and-pencil administration methods \citep{Bugbee1996}.

\section{Literature Review}
While substantial research has compared paper-and-pencil tests (PPT) with online computer-based tests (CBT) on graded, high-stakes assessments, little of it has focused on low-stakes RBAs as pretests and posttests in college settings, for which participation may be optional. In investigations of low-stakes assessments, it is critical to look at participation rates as well as performance results. If CBTs lead to lower participation rates or skewing of participation rates towards particular types of student, then using CBTs may lead to misleading or unusable data. If CBTs impact student performance on assessments, then comparisons to PPT data may be difficult or impossible to make. In our review of the literature we will examine what research shows about the impact on student participation rates and performance of transitioning assessments from PPTs to CBTs. 

\subsection{Participation rates}
To determine normative participation rates for RBAs and what factors are related to them, we reviewed 23 studies using RBAs in courses that were similar to those examined in our study (i.e., introductory physics courses). The studies we identified reported pretest and posttest results for either the FCI, the Force and Motion Conceptual Evaluation (FMCE) \citep{Thornton1998}, or the Brief Electricity and Magnetism Assessment (BEMA) \citep{Ding2006}. Of these 23 published studies, only four provided enough information about their data for us to evaluate the participation rates \citep{Author,Kost2009a,Kost-Smith2010a,Cahill2014}. Three provided sufficient data to compare participation rates across gender and course grade. Each of the four papers reported only their \textit{matched data} after performing listwise deletion. The studies reported that participation rates ranged from 49\% to 80\%, that female students were 5\% to 19\% more likely to participate, and that students who participated had higher grades than those that did not (see Table 1).

%Table I
\begin{table}[t]
\caption{Participation and GPA for students in previous studies. Participation rates varied across the studies, tended to be higher for female students, and higher for students with higher course grades.}
{\scriptsize
\begin{tabular}{lcccccccccc}  \hline\noalign{\smallskip}
Source&Gen.&\multicolumn{3}{c}{Part. Grade}&\multicolumn{3}{c}{Non-Part. Grade}&Part.&$\Delta_{grade}$&Odds\\
&&Mean&N&SD&Mean&N&SD&Rate&&F/M \\ \hline\noalign{\smallskip}
\multirow{2}{*}{Author} &M&2.69&90&1.28&2.1&92&1.28&0.49&0.59&\multirow{2}{*}{1.37}\\ %\makecell{Nissen et\\ al., 2016}
&F&2.78&27&1.26&2.05&13&1.16&0.68&0.73&\\ \hline\noalign{\smallskip}
\multirow{2}{*}{\makecell{Kost et\\ al., 2010}}&M&2.85&1257&0.8&1.93&500&1.1&0.72&0.92&\multirow{2}{*}{1.11}\\
&F&2.8&447&0.8&1.96&114&1.2&0.80&0.84&\\ \hline\noalign{\smallskip}
\multirow{2}{*}{\makecell{Kost et\\ al., 2009}}&M&2.82&1563&0.8&2.14&1152&1.2&0.58&0.68&\multirow{2}{*}{1.09}\\
&F&2.74&533&0.8&1.89&315&1.1&0.63&0.85&\\ \hline\noalign{\smallskip}
\multirow{4}{*}{\makecell{Cahill \\et al.,  \\2014}}&All&-&366&-&-&314&-&0.54&-&-\\
&All&-&773&-&-&448&-&0.63&-&-\\ 
&All&-&360&-&-&219&-&0.62&-&-\\
&All&-&738&-&-&384&-&0.66&-&-\\ \noalign{\smallskip} \hline
\end{tabular}
}
\end{table}

\par 	Because few studies have investigated student participation on low-stakes assessments in physics learning environments, we expanded our literature review to cover a wider range of fields. Research into student participation rates on low-stakes assessments has primarily focused on end-of-course and end-of-degree evaluations \citep{Dommeyer2004,Stowell2012, Bennett2010,Nulty2008,Nair2008,Goos2017}. All of these studies of participation rates examine non-proctored, low-stakes CBTs because high-stakes and proctored tests (e.g. course finals or the GREs) typically require participation. The majority of these studies examine how instructor or institutional practices affect overall student participation rates. These studies found that  reminders and incentives for participation increased overall participation rates. In an examination of end-of-course evaluations from over 3,000 courses, \citet{Goos2017} disaggregate overall participation rates to test for selection bias in students' participation. They found that there was a positive selection bias that had non-negligible effects on the average evaluation scores. While these studies did not use data from RBAs, they provide context for the instructor practices we examine and the analysis we perform in our research.

\par 	\citet{Bonham2008} was one of the first to examine student participation rates on RBAs. He examined data from college astronomy courses where assessments were administered both online outside of class as CBTs and in class as PPTs. Students completed a locally made concept inventory and a research-based attitudinal survey. The students (N=559) were randomly assigned to two assessment conditions with either the concept inventory done in-class and the attitudinal survey done outside of class via an online system or the reverse. \citet{Bonham2008} examined the impact of faculty practices on student participation rates by comparing student participation across classes that offered varying incentives to participate. Student participation rates on the CBTs were 8\% to 27\% lower than on the PPTs. Courses that offered more credit, reminders in class, and email reminders had higher student participation rates. 

\par	In preliminary work for this study, %\citet{Jariwala2016}
 \citet{Author} examined student participation rates on RBA pretests and posttests across several physics courses. The study included 693 students in three physics courses taught by five instructors at a large public university. Instructors used the Learning About STEM Student Outcomes (LASSO) platform to administer the CBTs. The LASSO platform is a free online system that hosts, administers, scores, and analyzes student pretest and posttest scores on science and math RBAs. The LASSO platform is described in detail in the methods section. The researchers employed an experimental design to randomly assign each student an RBA to complete in class on paper and an RBA to complete outside of class using LASSO. Average posttest participation rates for the five instructors ranged from 18\% to 90\% for CBTs and 55\% to 95\% for PPTs. While some instructors had significantly lower participation rates for CBTs than for PPTs, others had rates that were quite similar. Interviews of the faculty about their CBT administration practices found several commonalities between the courses with higher participation rates. Instructors with higher CBT participation rates gave their students credit for participating and reminded their students to complete the assessment both over email and during class. 

\par 	The general trends in findings for all the studies on participation rates were that participation rates on both PPT and CBT varied, and that there was the potential for skewing of data by student demographics and course grades. Participation rates for CBTs increased when instructors provided students with some form of credit for participating and with reminders to complete the survey. While all studies found similar results, most primarily relied on descriptive statistics to support their claims. The lack of statistical modeling in these publications means they lack precise claims, such as how much difference in participation rates is caused by giving students reminders or offering credit. The studies also largely ignored the impact of student demographics on participation rates.  For example, none of the studies examined how student gender or performance in a class impacted their likelihood of participating. These factors must be taken into account to make generalizable claims.

\subsection{Performance}
\par	Significant work has gone into examining the impact of CBT and PPT administration on student performance. Interest in the impact of CBTs picked up in the 1990s as testing companies (e.g., the Educational Testing Service and the College Board) transitioned services to computers and digital Learning Management Systems (e.g., Blackboard Learn and Desire2Learn) emerged as common course tools \citep{Bugbee1996}. These shifts in testing practices led to several studies into the impact of computerizing high-stakes, proctored assessments in both K-12 \citep{Kingston2008,Wang2007,Wang2007a} and university settings \citep{Prisacari2017,Candrlic2014,Wellman2004,Anakwe2008,Clariana2002}. Research across these settings generally found that performance on proctored computerized versions of high-stakes assessments was indistinguishable from performance on traditional PPTs. These studies make no claims whether their findings are generalizable to low-stakes RBAs.
\par	Only a handful of studies have examined the impact of computerized administration of low-stakes RBAs on university student performance. In Bonham's \citeyear{Bonham2008} research into college astronomy courses, he drew a matched sample from students who completed the in-class and outside-of-class surveys. He concluded that there was no significant difference between unproctored CBT and PPT data collection. However, examining Bonham's results reveals that there was a small but meaningful difference in the data. The results indicated that the online concept inventory scores were 6\% higher than the in-class scores on the posttest. For these data 6\% is an effect size of approximately 0.30. While this difference is small, lecture-based courses often have raw gains below 20\%; a 6\% difference would therefore skew comparisons between data collected with CBT and PPT assessment conditions. Therefore, the results of the study do not clearly show that low-stakes tests provide equivalent data when collected in class with PPTs or outside of class with CBTs. 
\par	In an examination of 136 university students' performance on a biology test and a biology motivation questionnaire, \citet{Chua2013} used a Solomon four-group experimental design to assess differences between tests administered as CBTs and PPTs. The participants were 136 undergraduate students in a teacher education program. The researchers created four groups of 34 students and assigned each to one of four assessment conditions: (1) PPT posttest, (2) PPT pretest and posttest, (3) CBT posttest, and (4) CBT pretest and posttest. The posttest was administered two weeks after the pretest. This design allowed the analysis to differentiate between differences caused by taking the pretest and differences caused by doing the test as a CBT instead of PPT. After accounting for the effects of taking the pretest, the researchers found no significant differences between the tests administered as CBTs and those administered as PPTs. While the study uses a strong experimental design, the sample size is small (N=34/group) which brings the reliability and generalizability of the study into question.
\par	\citet{Chuah2006} examined the impact of assessment conditions on student performance on a low-stakes personality test. They assigned the participants (N=728) to one of three assessments conditions: (1) PPT, (2) proctored CBT, and (3) unproctored CBT. They used mean comparison and Item Response Theory to examine participant performance at both the assessment and item levels. Their investigation found no meaningful differences in performance between the three assessment conditions. The authors concluded that their analysis supports the equivalence of CBTs and PPTs for personality tests.
\par	As described above, even among the studies that are most closely aligned with our research questions, very few of them directly examined how student responses on low-stakes, unproctored administration of CBTs compare to responses on PPTs. Those that have examined these issues tend to have small sample sizes and do not find consistent differences, making it difficult to support reliable and generalizable claims using their data.

\section{Research Questions}
\par	The purpose of the present study is to examine whether concept inventories and attitudinal surveys administered as low-stakes assessments online outside of class as CBTs provide equivalent data to those administered in class as PPTs. We examine equivalence between CBT and PPT administrations for both participation and performance.

To examine equivalence of participation, we ask the following three research questions:
\begin{enumerate}
\item How do instructor administration practices impact participation rates for low-stakes RBAs, if at all?
\item How are student course grades related to participation rates for low-stakes RBAs, if at all?
\item To what extent does participation differ across demographic groups?
\end{enumerate}
To examine equivalence of performance, we ask the following research question:
\begin{enumerate}
\setcounter{enumi}{3}
\item How does assessment condition (PPT vs CBT) impact student performance on low-stakes RBAs, if at all?
\end{enumerate}

If an online data collection platform can provide equivalent quantity and quality of data to paper-based administration, then the platform addresses many of the instructors' needs that \citet{Madsen2016} identified, and therefore lowers barriers for instructors to assess and transform their own courses. A second major benefit of the widespread use of an online data collection system like the LASSO platform is that they can aggregate, anonymize, and make all the data available for research (more details on the LASSO platform are provided in the Methods). The size and variety of this data set allows researchers to perform investigations that would be underpowered if conducted at only a few institutions or would lack generalizability if only conducted in a few courses at a single institution. 

\section{Methods}
\subsection{Setting}
The data collection for the study occurred at a large regional public university in the United States that is a Hispanic-Serving Institution (HSI) with an enrollment of approximately 34,000 undergraduate students and 5,000 graduate students. The university has a growing number of engineering majors and large numbers of biology and pre-health majors, all of whom are required to take introductory physics. 

\par We collected data from 27 sections of three different introductory physics courses (algebra-based mechanics, calculus-based mechanics, and calculus-based electricity \& magnetism [E\&M]) over two semesters (Table 2). Algebra-based mechanics was taught in sections of 80-100, without research-based instructional materials or required attendance. The calculus-based courses were taught in sections of 30-50, were supported by Learning Assistants (LAs), and used research-based instructional methods; incentives for attendance varied by instructor. In a typical semester, the Department of Physics offers four to six sections of each of these courses. We discarded data from 2 of the 27 sections due to instructor errors in administering the assessments. The data from the 25 sections analyzed in this study are described in Table 2.

%Table II
\begin{table}[t]
\caption{Course demographic data and instruments used.}
{\scriptsize
\begin{tabular}{lccccccccc} \hline \noalign{\smallskip}
&\multicolumn{4}{c}{\underline{Semester 1 (Spring 2016)}}&\multicolumn{4}{c}{ \underline{Semester 2 (Fall 2016)}}& \underline{Instruments}\\ \noalign{\smallskip}
&Sect.&Stud.&Male &URM &Sect.&Stud.&Male&URM&CI/AS\\ \hline\noalign{\smallskip}
A Mech&2&194&58\%&46\%&6&490&50\%&52\%&FCI/CLASS\\
C Mech&5&188&74\%&45\%&4&175&67\%&60\%&FCI/CLASS\\
C E\&M&4&117&70\%&52\%&4&146&74\%&47\%&CSEM/CLASS\\ \noalign{\smallskip}
Total&11&499&67\%&47\%&14&811&58\%&53\%&-\\ \noalign{\smallskip} \hline
\end{tabular}
}
\end{table}

\subsection{Design of the data collection}
\par	The study used a between-groups experimental design. We used stratified random sampling to create two groups within each course section with similar gender, race/ethnicity, and honors status makeups. The institution provided student demographic data. Group 1 completed a concept inventory (CI) online outside of class using the LASSO platform, and an attitudinal survey (AS) in class using paper and pencil (Figure 1). Group 2 completed the CI in class and the AS online outside of class. Within each course, both groups completed the in-class assessment at the same time and had the same window of time to complete the online assessment. Assessments were administered at the beginning and end of the semester. 

%figure1
\begin{figure}
\includegraphics[width=.7\columnwidth]{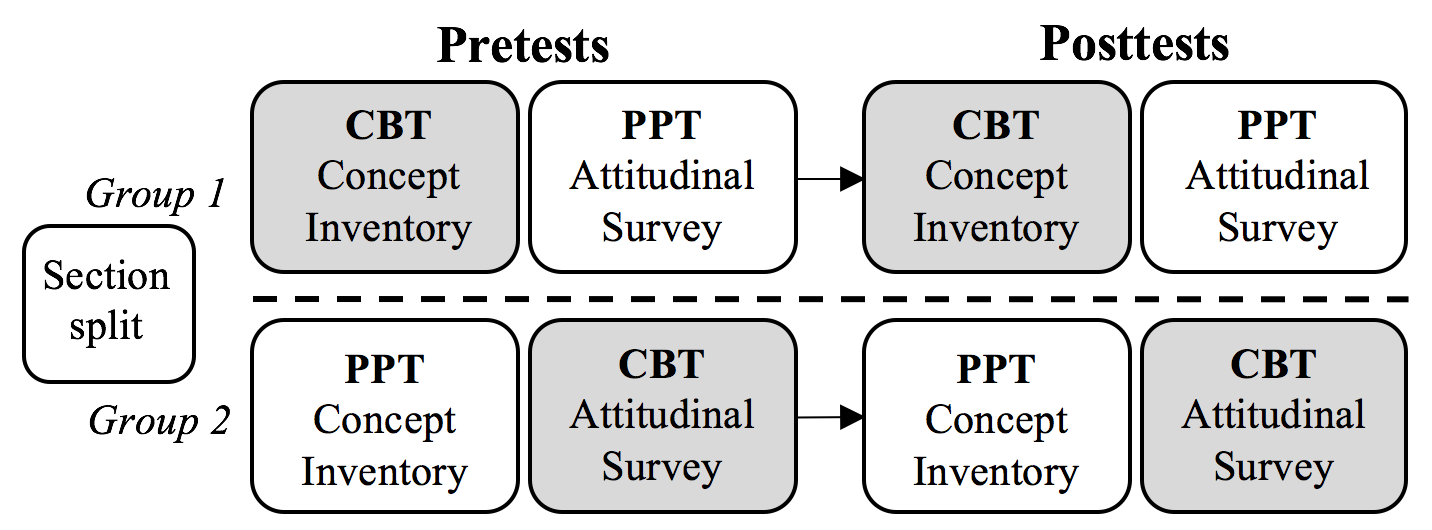}
\caption{Student groupings for RBA assignments using stratified random sampling. Each student takes one assessment online using LASSO and one in-class on paper at the beginning and again at the end of the semester.}
\end{figure}
\par	The LASSO platform (\url{learningassistantalliance.org/public/lasso.php}) hosts, administers, scores, and analyzes RBAs online. When setting up a course in LASSO, instructors answer a set of questions about their course, select their assessments, and upload a course roster with student emails. When instructors launch a pretest their students receive an email from the LASSO platform with directions on how to participate and a unique link that takes them to their assessment page. The first question students answer is whether they are over 18 years of age and are willing to have their data anonymized and made available to researchers. Students then complete a short set of demographic questions and begin their assessment. Instructors can track which students have participated in real-time and use the LASSO platform to generate reminder emails for students who have not yet completed the assessment. Near the end of the semester, faculty launch the posttest and the process of data collection repeats. After the posttest closes, instructors receive a report on their students' performance. Instructors can access all of their students' responses at any time. Data from participating courses are added to the LASSO database where they are anonymized, aggregated with similar courses, and made available to researchers with approved IRB protocols.
\par	Paper assessments were collected by the instructors, scanned using automated equipment, and uploaded to the LASSO platform, where the research team matched it with the CBT data collected directly through the platform. The research team downloaded the full set of student data from the LASSO platform and combined it with student course grades and demographic data provided by the institution. The data analysis did not include students who joined the class late or dropped/withdrew from the course because the research team could not assign them to a treatment group. Prior to applying filters to remove these students, the sample was 1,487 students. With these filters applied, the total sample was 1,310 students in 25 course sections. 	
\par	Students in both mechanics courses completed the 30 question Force Concept Inventory (FCI) \citep{Hestenes1992}. Students in the E\&M course completed the 32 question Conceptual Survey of Electricity and Magnetism (CSEM) \citep{Maloney2001}.  We scored both CIs on a 0-100\% scale. Students in all the courses completed the same AS, the Colorado Learning Attitudes about Science Survey (CLASS). The CLASS measures eight separate categories of student beliefs compiled from student responses to 42 questions. Responses are coded as favorable, neutral, or unfavorable based on agreement with expert responses. We analyzed the overall favorable score in the present study on a 0-100\% scale. We obtained course grades from the course instructors and student demographics from the institution. 
\par 	During the first semester of data collection \citep{Authors}%\citep{Jariwala2016}
, the research team provided the instructors with little guidance on how to motivate students to complete their CBT. Participation rates varied greatly across instructors. The research team asked the instructors what practices they used to motivate students, and identified four instructor practices associated with higher student CBT participation rates. The research team adopted these four instructor practices as recommended practices:

\begin{enumerate}
\item multiple email reminders, 
\item multiple in-class announcements,
\item participation credit for the pretest, and
\item participation credit for the posttest.
\end{enumerate}

\par 	During the second semester of data collection, the research team advised all instructors to use the recommended practices to increase student participation. At the end of the second semester, we asked the instructors what they had done to motivate students to participate in their CBTs. We used instructor responses to assign each section a Recommended Practices score ranging from zero to four according to the number of recommended practices they implemented. All analyses presented in this article include both semesters of data.

\section{Analysis}
\par 	We used the HLM 7 software package to analyze the data using Hierarchical Linear Models (HLM). HLM is a method of modeling that leverages information in the structure of nested data. In our data, measurements (student scores on assessments) nested within students and students nested within course sections, as shown in Figure 2. HLM also corrects for the dependencies created in nested data \citep{Raudenbush2002}. These dependencies violate the assumptions of normal Ordinary Least Squares regression that each measure is independent of each other, an assumption which is not met when comparing students grouped in different classes. HLM can account for these interdependencies by allowing for classroom-level dependencies. In effect, HLM creates unique equations for each classroom and then uses those classroom-level equations to model an effect estimate across all classrooms. Within the HLM 7 software, we used the hypothesis testing function to generate means and standard errors from the models for plots and comparisons.

\begin{figure}
\includegraphics[width=.7\columnwidth]{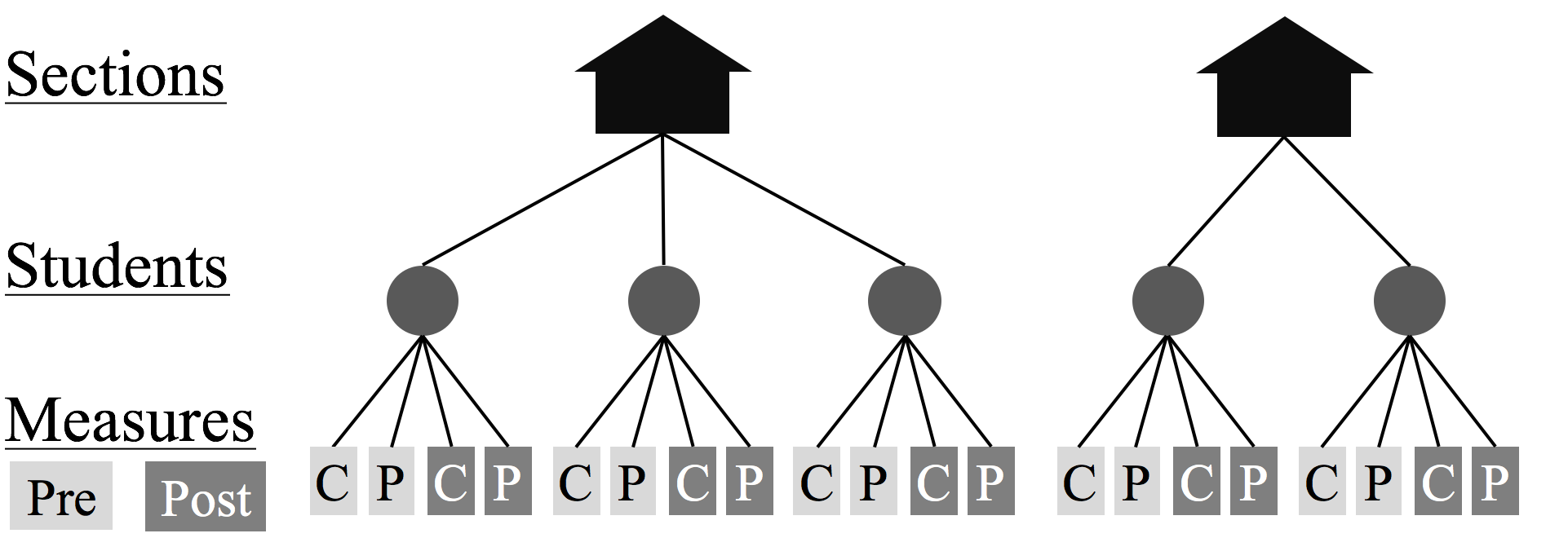}
\caption{The structure of the data is hierarchical with measures (either participation or scores) nested in students nested in course sections.}
\end{figure}

\par We investigated the performance research questions with one set of HLM models and the participation research questions with a separate set of Hierarchical Generalized Linear Models (HGLM). The two different types of HLM were necessary because the outcome variable was binary in the participation models (students did or did not participate) and continuous in the performance models (RBA score).
\par	For both the participation and performance models we built each model in several steps by adding variables. We compared both the variance and the coefficients for each model. Comparing the total variance in each of the models informed the strength of the relationship between the variables in the model and students participation. For example, variables that related to participation would reduce the total variance in the models that included them. The more that the variance reduced the stronger the relationship between the variables and participation. In HLM, the variance is also distributed across the levels of the model: our 3-level models measure variance within students, between students within a section, and between sections. We are interested in both the change in the total variance and the change at specific levels when variables are added. For example, when we add the section level variables to the models such as course type or instructor practices, we are interested in how much the variance between sections is reduced. The model coefficients size indicated the strength of the relationship between each variable and the outcome variable. Together, variance and coefficient size allow us to identify the extent to which the variables of interest predict student participation and performance.

\subsection{Participation}
To investigate students' participation rates in the computer versus paper-and-pencil assessments, we differentiated between each assessment by assessment condition and assessment timing using four dummy variables: pre-CBT, post-CBT, pre-PPT and post-PPT. Our preliminary HGLM analyses indicated that there was no difference in participation between the AS and CI instruments, so to keep our models concise we did not include variables for instruments in the models we present. We built an HGLM of students' participation rates for the PPT and the CBT on both the pretest and posttest. The HGLM was a population-averaged logistic regression model using Penalized Quasi-Likelihood (PQL) estimation because the outcome variable was binary (whether or not students completed the assessment). We used PQL because it was easily available in the HLM software and less computationally intensive than other estimation techniques. However, PQL overestimates the probability of highly likely events \citep{Capanu2013}. To address this concern, we compared the 3-level HGLM models we report in this article to four 2-level HGLM models that used adaptive Gaussian quadrature estimation. There were no meaningful differences in the models or the inferences that we would make from the models. For simplicity, we only report the three-level HGLM model that used full PQL estimation. 
\par	The data are nested in three levels (Figure 2): the four measures of participation nested within students, and the students nested within course sections. The outcome variable for these models was whether students had participated in the assessment (0 or 1). In the final model (Equation 1-7), we included dummy variables for the four assessment condition and timings (CBT pre, PPT pre, CBT post, and PPT post) at level 1, students' final grades in the course as four dummy variables (0 or 1 for each of the grades A, B, C, and D) at level 2, gender (male = 0 and female = 1) at level 2, and a continuous variable for recommended practices (0 to 4) at level 3. The structure of these variables is laid out in Table 3. The dummy variable for an F grade is not included in the equation because it is integrated into the intercept value. The models did not include the recommended practices for the PPTs because the practices focused on improving participation on the CBTs. The value of the recommended practices variable was the cumulative number (0 to 4) of recommended practices that faculty used to motivate their students to participate in the CBTs. The models included students' grades in the course because analysis of the raw data showed that students' course grades positively related to participation; we included course grades as dummy variables rather than as a continuous variable because there was a non-linear relationship between course grade and participation. Our preliminary analysis also included a dummy variable for race/ethnicity but we did not include it in the final model because it was not predictive of student participation.  
\par	In a logistic model, the coefficients for the predictors are logits ($\eta$), or logarithms of the odds ratio. We generated probabilities for different groups of students participating by using the model to create a logit for that probability and then converting the logit to a probability using Equation 8.

%Table III
\begin{table}[t]
\caption{Variables used in the final participation and performance models (outcome variables in bold).}
{\scriptsize
\begin{tabular}{cccc} \hline\noalign{\smallskip}
Model&\multirow{2}{*}{Structure}&\multicolumn{2}{c}{Variables}\\ \cline{3-4}\noalign{\smallskip}
Level&&Participation&Performance\\ \noalign{\smallskip}\hline\noalign{\smallskip}
\multirow{2}{*}{1}&\multirow{2}{*}{Assessment}&\textbf{Participation (0 or 1)}& \textbf{Score 0\% to 100\%} \\
&& Assessment condition and timing&Assessment timing\\ \noalign{\smallskip}\hline\noalign{\smallskip}
\multirow{2}{*}{2}&\multirow{2}{*}{Students}&Course Grade& \multirow{2}{*}{Assessment condition}\\
&&Gender&\\ \noalign{\smallskip}\hline\noalign{\smallskip}
3&Sections&Recommended Practices (CBT only)&Course Type\\
\noalign{\smallskip}\hline
\end{tabular}
}
\end{table}

Level-1 Equations
\begin{align}
Probability(Participation_{ijk} = 1|\pi_{jk}) = \phi_{ijk}\\
log[\phi_{ijk}/(1-\phi_{ijk})]=\eta_{ijk} 
\end{align}
\begin{multline}
\eta_{ijk}=\pi_{1jk}*CBT PRE_{ijk}+\pi_{2jk}*CBT POST_{ijk} \\+ \pi_{3jk}*PPT PRE_{ijk}+\pi_{4jk}*PPT POST_{ijk}
\end{multline}
\par Level-2 Equations. There are 4 level-2 equations, one for each $\pi$.
\begin{multline}
\pi_{ijk} = \beta_{i0k} + \beta_{i1k}*Gender_{jk} + \beta_{i2k}*A_{jk} + \beta_{i3k}*B_{jk} \\+ \beta_{i4k}*C_{jk} + \beta_{i5k}*D_{jk} + r_{jk}
\end{multline}
\par Level-3 Equations. There are 24 level-3 equations, 2 include a variable for practices, 22 do not and are illustrated by Equation 7.
\begin{align}
\beta_{10k}=\gamma_{100}+\gamma_{101}*Practices_k+ u_{1jk}\\
\beta_{20k}=\gamma_{200}+\gamma_{201}*Practices_k+ u_{2jk}\\
\beta_{ijk}=\gamma_{ij0}+ u_{ijk}
\end{align}

\begin{equation}
\phi=10^\eta/(1+10^\eta)
\end{equation}

\par	We built the model in three steps: (1) differentiating between the pretest and posttest for the CBT and PPT assessment conditions, (2) adding the level 3 predictor for the number of recommended practices the instructor used, (3) adding level 2 predictor for course grade and gender. On their own, the effect that the different model coefficients have on participation rates is difficult to interpret because they are expressed in logits. Part of the difficulty is that the size of each coefficient cannot be directly compared because the effect of a coefficient on the probability of participation depends on the other coefficients to which it is being added (e.g., the intercept). For example, a logit of 0 is a 50\% probability, 1 is approximately 90\%, and 2 is 99\%. Thus, a 1.0 shift in logits from 0 to 1 is a much larger change in probability than the 1.0 shift from 1 to 2 logits. The importance of the starting point was particularly salient for interpreting the coefficients in our HGLM models because the intercepts for the pre/post assessment conditions varied from a low of -2.7 to a high of 2.3. To simplify interpreting the results of the model, we used the hypothesis testing function in the HLM software to generate predicted logits and standard errors for each of the combinations of variables and converted the logits to probabilities with error bars of one standard error. In our analyses we focused on posttest participation rates because they are the more limiting rates for data collection, and because the posttests contain information about the effects of the course whereas the pretests only contain information about the students who enroll in the course. 
\par	Our investigation of differences in participation rate by course grade and gender used other analyses in addition to the coefficients and variance output by the HGLM model. For comparing the differences in participation rates by gender we used the odds ratio, which the HGLM produces as an output and which are easily calculated for studies in the published literature. An odds ratio of 1.0 indicates that male and female students were equally likely to participate. An odds ratio greater than 1.0 indicates that female students were more likely to participate than male students. If the confidence interval for the odds ratio includes 1.0, then it is not statistically significant. Comparing the differences in participation by course grade was more difficult because the HGLM does not produce an output that is comparable to the mean grades for participants and nonparticipants, which is the statistic that prior studies report. Therefore, we also reported these raw statistics to situate our study within the existing literature.

\subsection{Performance}
To investigate differences in performance between tests administered as CBTs and PPTs (Research Question 4), we built separate HLM models for the CI and AS scores. It was possible to combine these models into a single multivariate HLM. However, multivariate HLMs are more complex to both analyze and report and the HLM software documentation recommends that researchers start with separate models for each variable \citep{Raudenbush2011}. After producing our models we concluded that the two models were sufficient for our purposes. The HLM performance models for the CI and AS data had identical structures. All performance models used RBA score as the outcome variable. The models included a level-1 variable (post) to differentiate between the pretest and posttest. The variable of interest for the models that addressed Research Question 4 was assessment condition at level 2. We also included predictor variables at level 3 for each of the three courses because performance varied across the course populations, and it allowed us to make comparisons of the effect of assessment condition across the multiple courses for both the pretest and posttest. These comparisons had the advantage of  indicating whether there was a consistent difference in scores (e.g., CBT was always higher), even if that difference was too small to be statistically significant. Initially, we included level-2 variables to control for course grade, gender, and race/ethnicity because these variables relate to performance on RBAs \citep{Madsen2013,Author}.%Vandusen2015
However, these demographic variables had no effect on the impact of assessment conditions on student performance in our models. For brevity, we excluded these variables from the models we present here. 
\par	The final performance model included RBA score as the outcome variable and predictor variables for posttest (level 1), assessment condition (level 2), and course (level 3) (Equation 3). The variables used in the final model are shown in Table 3. We built the model in three steps: (1) a level-1 variable for posttest, (2) then add a level-2 variable for assessment condition, and (3) add level-3 variables for each course. To determine how much variance in the data was explained by each of the variables, we compared the total variance between each of the models. The reduction in the variance between the models indicated the strength of the relationship between the variables and performance by showing how much information about performance the added variables provided. For example, if there were large differences in performance between PPTs and CBTs, then the addition of CBT to the model would decrease the total variance. One distinction between HLM and OLS regression is that in OLS additional variables always reduces the unexplained variance, whereas in HLM the variance can increase if a nonsignificant variable is added to the model \citep[p.~150]{Raudenbush2002}. We used the hypothesis testing function in the HLM software to generate predicted values and standard error for each of the courses' pretest and posttest scores, for both assessment conditions, to inform the size and reliability of any differences between assessment conditions.
\par	For the performance analyses, we replaced missing data using Hierarchical Multiple Imputation (HMI) with the mice package in R. We discuss the rate of missing data in the Results Participation section (7.1) below. Multiple Imputation (MI) addresses missing data by (1) imputing the missing data \textit{m} times to create \textit{m} complete data sets, (2) analyzing each data set independently, and (3) combining the \textit{m} results using standardized methods \citep{Dong2013}. MI is preferable to listwise deletion because it maximizes the statistical power of the study \citep{Dong2013} and has the same basic assumptions. HMI is MI that takes into account students being nested in different courses and that their performance may be related to the course they were in. Our HMI produced \textit{m}=10 complete data sets. In addition to pretest and posttest scores, the HMI included variables for course, course grade, gender, and race/ethnicity. We used the HLM software to automatically run analyses on the HMI datasets.

Level-1 Equations
\begin{equation}
SCORE_{ijk}=\pi_{0jk} + \pi_{1jk}*POST_{ijk}+ e_{ijk}
\end{equation}
\par Level-2 Equations
\begin{align}
\pi_{0jk}=\beta_{00k} + \beta_{01k}*Condition_{jk}+r_{0jk}\\
\pi_{1jk}=\beta_{10k} + \beta_{11k}*Condition_{jk}
\end{align}
\par Level-3 Equations
\begin{align}
\beta_{00k}= \gamma_{001}*AlgMech_k+\gamma_{002}*CalcMech_k+\gamma_{003}*CalcE\&M_k +u_{00k}\\
\beta_{01k}=\gamma_{011}*AlgMech_k+\gamma_{012}*CalcMech_k+\gamma_{013}*CalcE\&M_k+u_{01k}\\
\beta_{10k}=\gamma_{101}*AlgMech_k+\gamma_{102}*CalcMech_k+\gamma_{103}*CalcE\&M_k+u_{10k}\\
\beta_{11k}=\gamma_{111}*AlgMech_k+\gamma_{112}*CalcMech_k+\gamma_{113}*CalcE\&M_k+u_{11k}
\end{align}

\section{Results}
First, we present the results for the participation analysis. These results include descriptive statistics and the HGLM models. Then we present the results for the performance analysis.
\subsection{Participation}
We first compare the raw participation rates for the CBTs and PPTs -- overall, by gender, and by grade -- to participation rates reported in prior studies. This comparison identifies the extent to which participation in this study was similar to participation in prior studies and informs the generalizability of our findings. Prior studies report grade and gender differences in participation in aggregate so we cannot compare their findings to our HGLM outputs, which differentiate between each course grade. Therefore, we compare the raw differences in mean course grades for participating and nonparticipating students in our data to the differences reported in prior studies.
\par	Following our comparison of the raw data we present three HGLM models.  Model 1 differentiates between the pretests and posttests for the two assessment condition (CBT and PPT). The second model addresses Research Question 1 by accounting for how instructor use of the recommended practices related to student participation. Model 3 addresses Research Questions 2 and 3 by including variables for student gender and course grade.

\subsubsection{Descriptive statistics}
The descriptive statistics show that the overall PPT participation rate is higher than the overall CBT participation rate for pre and post administration of both the CI and AS, as shown in Table 4. These raw participation rates do not account for differences in participation across course sections. These rates all fall within the range found in prior studies shown in Table 1. Gender differences in participation in the raw data for this study are small and are smaller than those reported in prior studies. Differences in course grades between those that did and did not participate are large and are similar in size to those reported in prior studies. However, these comparisons between the present study and prior studies are only approximations. The prior studies reported matched data and in some of these studies it is unclear if they included all students who enrolled in the course, only students who received grades, or only students who took the pretest. The present study includes only students who enrolled in the course prior to the first day of instruction and who received a grade in the course. While these differences between the present study and prior studies make it difficult to compare participation rates, the approximate comparison indicates that the present study is not outside the boundaries of what researchers have reported in prior studies. 
 
%Table IV
\begin{table}[t]
\caption{ Participation rates for pre and post CBT and PPT administered exams comparing participation for the type of instrument, the gender of the participants, and the grades of the participants and nonparticipants.}
{\scriptsize
\begin{tabular}{lccccccccc} \hline\noalign{\smallskip}
&&& & \multicolumn{3}{c}{\underline{Gender Differences}}&\multicolumn{3}{c}{\underline{Mean Course Grade}}\\\noalign{\smallskip}
Cond. &Time&AS&CI & Male&Female&Odds&Part. &Non-part.&$\Delta$\\ 
&& & &N=803&N=507&(F/M)& &&\\ \hline\noalign{\smallskip}
\multirow{2}{*}{CBT}&Pre&71\%&67\%&66\%&76\%&0.99&2.86&2.13&0.73\\
&Post&59\%&54\%&54\%&61\%&1.13&2.97&2.20&0.77\\
\multirow{2}{*}{PPT}&Pre&94\%&94\%&94\%&95\%&1&2.68&1.95&0.73\\
&Post&75\%&74\%&74\%&75\%&1&2.87&1.95&0.92\\
\noalign{\smallskip} \hline
\end{tabular}
}
\end{table}

\subsubsection{The relationship between participation and instructor practices}
After converting the logits given in Table 5 to probabilities, Model 1 shows participation rates of 83\% for the CBT pretest, 66\% for the CBT posttest, 100\% for the PPT pretest, and 95\% for the PPT posttest. These participation rates all exceed those calculated with raw data, a known issue with HGLM models as discussed in the Methods Section. Model 2 includes a variable for the number of recommended practices the instructors used in each section for the CBT pretest and posttests. Including recommended practices did not reduce the variance within assessments or between assessments within students (level 1 and level 2) for any of the assessment conditions, but it did explain a large part of the variance between sections for the CBT pretest and posttest, as shown in the bottom of Table 5. The variance in Model 2  is 15\% lower (from 0.820 to 0.700) for CBT pretests and 45\% lower (from 1.220 to 0.670) for the CBT posttests than in Model 1. This large decrease in variance indicates that the number of recommended practices instructors used to motivate their students to participate accounted for a large proportion of the difference in participation rates between sections on the assessments administered as CBTs.

%Table V
\begin{table}[t]
\caption{HGLM outputs for models comparing student participation on the CBT and PPT pretest and posttests by recommended practices (level 3), gender (level 2), and course grade (level 2).
}

{\scriptsize

\begin{tabular}{lcccccccc} \hline \noalign{\smallskip}
\multicolumn{9}{c}{Final estimation of fixed effects with robust standard errors}\\
&\multicolumn{2}{c}{Model 1}&&\multicolumn{2}{c}{Model 2}&&\multicolumn{2}{c}{Model 3}\\
&Coef.&\textit{p}&&Coef.&\textit{p}&&Coef.&\textit{p}\\
\multicolumn{9}{c}{CBT Pre $\pi_{1} $}\\ \hline\noalign{\smallskip}
\multicolumn{9}{l}{For Intercept 2 $\beta_{10}$}\\
    ~~For Int. 3 $\gamma_{100}$&0.679&\textless0.001&&0.256&0.434&&-0.671&0.081\\
    ~~Practices $\gamma_{101}$&-&-&&0.214&0.116&&0.182&0.097\\
  Gender $\gamma_{110}$&-&-&&-&-&&0.269&0.086\\
  D Grade $\gamma_{120}$&-&-&&-&-&&0.142&0.699\\
  C Grade $\gamma_{130}$&-&-&&-&-&&0.558&0.077\\
  B Grade $\gamma_{140}$&-&-&&-&-&&1.102&0.002\\
  A Grade $\gamma_{150}$&-&-&&-&-&&1.526&\textless0.001\\
\multicolumn{9}{c}{CBT Post $\pi_{2} $}\\ \hline\noalign{\smallskip}
 \multicolumn{9}{l}{For Intercept 2 $\beta_{20}$}\\
    ~~For Int. 3 $\gamma_{200}$ &0.296&0.118&&-0.767&0.008&&-2.678&\textless0.001\\
    ~~Practices $\gamma_{201}$ &-&-&&0.534&\textless0.001&&0.573&\textless0.001\\
  Gender $\gamma_{210}$&-&-&&-&-&&0.207&0.281\\
  D Grade $\gamma_{220}$&-&-&&-&-&&0.946&0.013\\
  C Grade $\gamma_{230}$&-&-&&-&-&&1.395&\textless0.001\\
  B Grade $\gamma_{240}$&-&-&&-&-&&2.057&\textless0.001\\
  A Grade $\gamma_{250}$&-&-&&-&-&&2.390&\textless0.001\\
\multicolumn{9}{c}{PPT Pre $\pi_{3} $}\\ \hline\noalign{\smallskip}
 \multicolumn{9}{l}{ For Intercept 2 $\beta_{30}$}\\
    ~~For Int. 3 $\gamma_{300}$ &2.290&\textless0.001&&2.266&\textless0.001&&1.361&\textless0.001\\
  Gender $\gamma_{310}$&-&-&&-&-&&0.130&0.619\\
  D Grade $\gamma_{320}$&-&-&&-&-&&0.496&0.281\\
  C Grade $\gamma_{330}$&-&-&&-&-&&0.675&0.062\\
  B Grade $\gamma_{340}$&-&-&&-&-&&0.835&0.042\\
  A Grade $\gamma_{350}$&-&-&&-&-&&0.909&0.049\\
\multicolumn{9}{c}{PPT Post $\pi_{4} $}\\ \hline\noalign{\smallskip}
  \multicolumn{9}{l}{For Intercept 2 $\beta_{40}$}\\
    ~~For Int. 3 $\gamma_{400}$&1.235&\textless0.001&&1.235&\textless0.001&&-0.706&0.047\\
  Gender $\gamma_{410}$&-&-&&-&-&&0.224&0.180\\
  D Grade $\gamma_{420}$&-&-&&-&-&&1.522&0.001\\
  C Grade $\gamma_{430}$&-&-&&-&-&&1.514&\textless0.001\\
  B Grade $\gamma_{440}$&-&-&&-&-&&2.166&\textless0.001\\
  A Grade $\gamma_{450}$&-&-&&-&-&&2.493&\textless0.001\\ \hline \noalign{\smallskip}
&&&&&&&&\\ 
\multicolumn{9}{c}{Level-1 and Level-2 Variance}\\ \hline\noalign{\smallskip}
CBT Pre $r_{1}$&\multicolumn{2}{c}{1.080}&&\multicolumn{2}{c}{1.080}&&\multicolumn{2}{c}{0.805}\\
CBT Post $r_{2}$&\multicolumn{2}{c}{1.170}&&\multicolumn{2}{c}{1.390}&&\multicolumn{2}{c}{1.077}\\
PPT Pre $r_{3}$&\multicolumn{2}{c}{1.130}&&\multicolumn{2}{c}{1.440}&&\multicolumn{2}{c}{1.156}\\
PPT Post $r_{4}$&\multicolumn{2}{c}{1.100}&&\multicolumn{2}{c}{1.200}&&\multicolumn{2}{c}{0.889}\\ \hline 
&&&&&&&&\\ 
\multicolumn{9}{c}{Level-3 Variance}\\ \hline\noalign{\smallskip}
CBT Pre $u_{10}$ &\multicolumn{2}{c}{0.820}&&\multicolumn{2}{c}{0.700}&&\multicolumn{2}{c}{0.740}\\
CBT Post $u_{20}$&\multicolumn{2}{c}{1.220}&&\multicolumn{2}{c}{0.670}&&\multicolumn{2}{c}{0.830}\\
PPT Pre $u_{30}$&\multicolumn{2}{c}{0.560}&&\multicolumn{2}{c}{0.485}&&\multicolumn{2}{c}{0.690}\\
PPT Post $u_{40}$&\multicolumn{2}{c}{1.340}&&\multicolumn{2}{c}{1.370}&&\multicolumn{2}{c}{1.180}\\
\noalign{\smallskip} \hline
\end{tabular}
}
\end{table}

\par	Using Model 2 we calculated the predicted participation rates for students on PPTs and CBTs in courses that used different numbers of recommended practices. We calculated the probabilities shown in the graph from the logits and standard errors calculated with the hypothesis testing function in the HLM software. The logit itself is easily calculated from the model. For example, the logit for CBT posttest participation in a course using 3 recommended practices is $ \eta = -0.767 + 3(0.534) = 0.834$. Using Equation 8 this logit gives a probability of 87\%. We then plotted these values and their error bars (1 standard error) in Figure 3.

\par	Figure 3 shows that when instructors used none of the recommended practices CBT participation rates were much lower than the PPT rates. When faculty used all four of the recommended practices, however, CBT participation rates matched PPT rates. All the predicted participation rates in these cases exceed 90\%. This participation rate is likely an overestimate caused by high probability predictions in HGLM using PQL. The model, however, is likely overestimating all the participation rates by a similar amount. For example, the predicted participation rates for a CBT posttest in a course using 4 recommended practices (96\%) and the PPT posttest (95\%) are effectively the same, so any overestimation in them should be the same.

\begin{figure}
\includegraphics[width=.7\columnwidth]{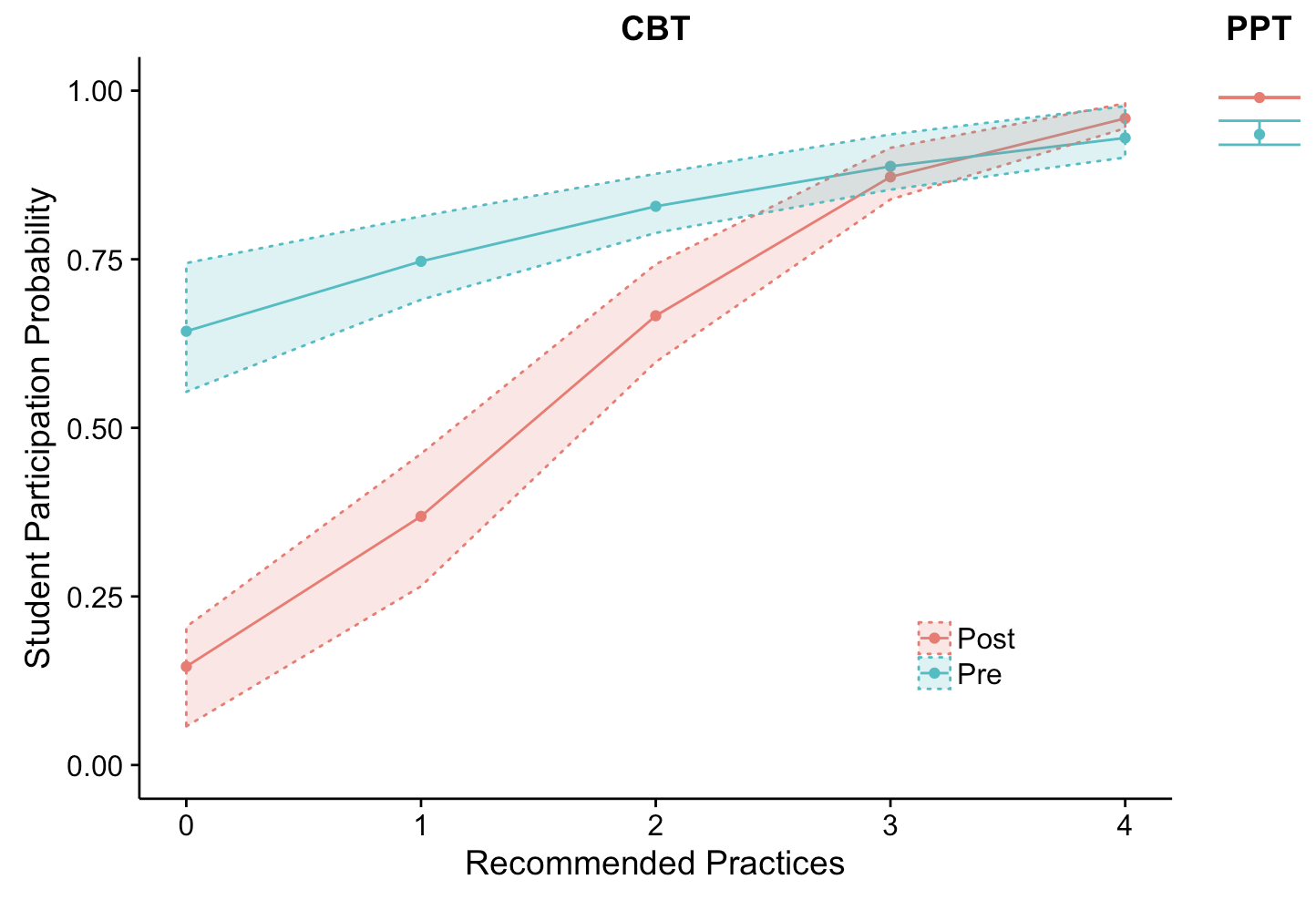}
\caption{Student predicted participation rates (+/- 1 standard error) by assessment condition and use of CBT recommended practices from Model 2. When instructors used all four recommended practices participation rates on CBT and PPT were similar.}
\end{figure}

\subsubsection{Participation by course grade and gender}
Model 3 includes variables for student gender and course grade. The addition of these variables decreased the variance between assessments as well as the variance within assessments between students for all CBT and PPT pretests and posttests by 20\% to 26\% (for example from 1.080 to 0.805) from Model 2. These variables tended to increase the variance between sections for Model 3 compared to Model 2 (+42\% to -14\%, for example from 0.485 to 0.690), indicating that there was unaccounted-for variation in how course grade and gender differentially related to participation in the different course sections. 
\par	Gender differences in participation in Model 3 were not statistically significant. However, all of the coefficients in the model indicated that female students were more likely to participate than male students. This higher participation rate for female students was also reported in all three of the prior studies. Therefore, it is possible that this is a real effect that is simply to small for our complex statistical model to identify as statistically significant. The odds ratios with 95\% confidence intervals comparing female to male participation rates calculated by the HLM software were 1.31 [0.96, 1.79] for the CBT pretest, 1.23 [0.84, 1.81] for the CBT posttest, 1.14 [0.67, 1.95] for the PPT pretest, and 1.25 [0.90, 1.75] for the PPT posttest. These odds ratios all predict higher female participation but have confidence intervals that include the value 1, indicating the difference in participation rates by gender was not statistically significant. These odds ratios, however, all fall within the range of odds ratios found in the three prior studies, which indicates the differences in participation rate by gender may be a consistent but small effect.
\par	We included student course grades as dummy variables (rather than as a single continuous variable) in Model 3 because our preliminary models indicated that the difference in participation between each grade was not linear. This nonlinear relationship is observable in the values in Model 3. For example, on the PPT posttest, the difference between students who earned Fs and Ds was 1.52 logits, whereas for students who earned As and Bs the difference was only 0.33 logits. In a linear relationship, it would have been approximately the same difference in logits between each adjacent pair of course grades. Entering the grades as four separate variables has the downside of complicating the model; however, these models more accurately portray the differences in participation between each of the course grades.
\par	Using the hypothesis testing function in the HLM software and Model 3, we generated the logits and standard errors for participation for each course grade under each assessment condition using the population mean for gender (0.39) and plotted these values in Figure 3. We used the mean value for gender so that we could focus on the differences in predicted average participation rates across assessment conditions and course grades. The figure does not include the PPT pretest because the model predicted that the participation rates across all course grades ranged from 96\% to 100\%, which is too small of a difference to be visible in Figure 4. Model 3 indicates that for both the CBT and PPT posttests all four grades (A-D) were statistically significantly more likely to participate than students who received an F in the course. Receiving a grade of F is not shown in the model because it is incorporated in the intercept. Figure 3 illustrates that students who received an A, B, or C had more similar participation rates than students who received a D or F. This is particularly evident when the participation rates are higher, such as on the PPT posttest or on both CBT assessments when 3 or 4 recommended practices were used. These results indicate that the data collection in these courses disproportionately represented higher achieving students in both assessment conditions. Given that the raw participation rates and differences in grades between those that did and did not participate were both similar to those reported in prior studies, these results strongly suggest that data collection with low-stakes RBAs systematically over represents high achieving students, regardless of assessment administration method.

\begin{figure}
\includegraphics[width=1\columnwidth]{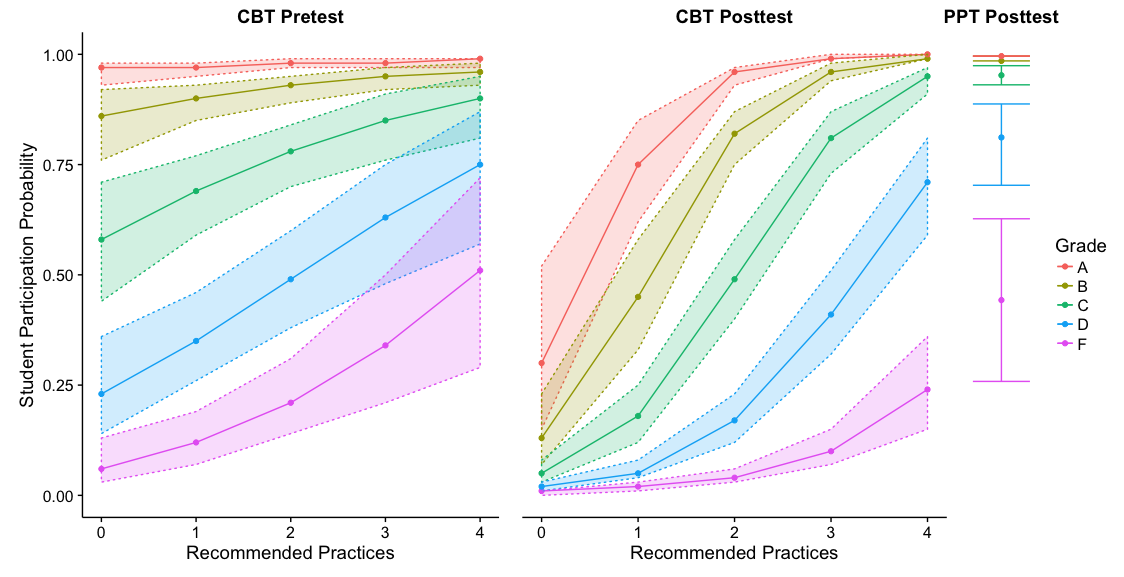}
\caption{Participation by course grade and recommended practices based on Model 3. Gender was entered into the linear equation as the mean value (0.39) to simplify the figure. PPT pretest is not shown because it's value varies from 96\% to 100\% across the course grades. The results indicate that there were large differences in participation across the different course grades and that when instructors used all four recommended practices rates on the CBT posttests were similar to participation rates on the PPT posttests across the different courses grades. }
\end{figure}

\subsubsection{Performance differences between RBAs administered as CBTs and PPTs}
As discussed in the Analysis Section, we built separate sets of models for performance on the concept inventories(CI) and on the attitudinal surveys (AS). We built these models in the same three steps to investigate performance differences between CBT and PPTs. The first model differentiated between pretest and posttest scores with a variable at level 1. The second model differentiates between assessments administered as CBTs or PPTs at with a variable at level 2. The third model added variables to differentiate between the three courses at level 3. In our analysis of these models, we first present the change in the variance between the models to identify how much of the variability in scores was explained by whether students took the assessments as CBT or PPT. Following the analysis of the variance, we present the size and consistency of the differences in scores between the two assessment conditions.
\par	For both the CI models and the AS models the total variance did not meaningfully decrease between Models 1, 2, and 3 (see bottom of Table 6). Model 2 differentiates between students who took the CI as PPT or CBT. For both the AS and CI models this differentiation caused the total variance in the models to increase. The increase in the total variance was very small for the CI models (\textless+1\% from 270.8 to 272.7) and small for the AS models (+2.8\% from 195.52 to 200.92). Increases in the variance for each sets of models is a strong indication that there were no differences in scores between those administered as CBTs and those administered as PPTs. Increases in the variance for both sets of models emphasizes that the tests provided equivalent data. However, it was possible that there were differences in some of the courses but not in others. To address this possibility, we developed Model 3 to compare CBT and PPT while differentiating between the three courses in the study. The total variance in Model 3 slightly decreased compared to Model 1 for the CI models (-1.7\%) and slightly increased for the AS models (+1.4\%). Given the shifts in variances' small sizes and disagreements in direction, the change in variance between the three models indicates that student performance on each assessment was equivalent whether administered as CBT or PPT. 
\par	CI Model 1 indicates that the average CI pretest score for all students was 31\% and that on average students gained 13\%. In Model 2, we differentiated between assessments administered as PPT or CBT. Model 2 for both CI and AS indicated that the differences in scores between PPTs and CBTs were very small and that these differences were not statistically significant. Specifically, on the pretest CBT scores were slightly higher than PPT scores (\textless+1\%) and on the posttest CBT scores were slightly lower than PPT scores (\textless-1\%). In Model 3, we disaggregated the data between the three course types, which also allowed us to differentiate between CI instruments. For the CI Model 3 there were substantial differences between the three courses. For the AS Model 3 there were small differences between the three courses. For both the AS Model 3 and the CI Model 3, the CBT condition was not a statistically significant predictor of score in any course. None of the assessment condition coefficients were statistically significantly different from zero on either the pretest or posttest. The hypothesis testing function in the HLM software generated means and standard errors based on the CI and AS Model 3s, presented in Figure 5. Figure 5 and both Model 3s all show that the differences between CBT and PPT scores were small (ranging from -2.1\% to 2.2\%) and that scores were not consistently higher in either assessment condition than in the other. In seven cases, the PPT was higher. In five cases, the CBT was higher. These results indicate that there was not a consistent, meaningful, or reliable difference in scores between assessments administered as CBTs and those administered as PPTs.

%Table VI
\begin{table}[t]
\caption{HLM outputs for models comparing performance between assessments administered as PPT and CBT for both the CI and AS. The models indicated that performance on the two modes of assessment was similar.
}
{\scriptsize
\begin{tabular}{lccccccc} \hline \noalign{\smallskip}
\multicolumn{8}{c}{Final estimation of fixed effects with robust standard errors}\\
&\multicolumn{3}{c}{CI Models}&&\multicolumn{3}{c}{AS Models}\\ \cline{2-4} \cline{6-8} \noalign{\smallskip}
&CI 1&CI 2&CI 3&&AS 1&AS 2&AS 3\\
&Coef. &Coef. &Coef. &&Coef. &Coef. &Coef. \\  \hline \noalign{\smallskip}
\multicolumn{8}{l}{For Intercept 1}\\ 
\multicolumn{8}{l}{~~For Intercept 2}\\
    ~~~~Intercept 3&30.99***&31.20***&-&& 43.97***&44.11***&-\\
    ~~~~Alg. Mech.&-&-&26.93***&&-&-&42.87***\\
    ~~~~Calc. Mech.&-&-&36.36***&&-&-&43.77***\\
    ~~~~Calc. E \& M&-&-&29.68***&&-&-&47.19***\\
\multicolumn{8}{l}{~~For Condition (CBT)}\\
    ~~~~Intercept 3&-&-0.42&&-&-&-0.25&-\\
    ~~~~Alg. Mech.&-&-&0.12&&-&-&-0.61\\
    ~~~~Calc. Mech.&-&-&-0.36&&-&-&1.63\\
    ~~~~Calc. E \& M&-&-&-1.18&&-&-&-2.21 \\
\multicolumn{8}{l}{For Post}\\
\multicolumn{8}{l}{~~For Intercept 2}\\
    ~~~~Intercept 3&13.04***&12.84***&-&&1.76**&1.33*&-\\
    ~~~~Alg. Mech.&-&-&7.45***&&-&-&1.66\\
    ~~~~Calc. Mech.&-&-&18.61***&&-&-&2.98*\\
    ~~~~Calc. E \& M&-&-&11.59***&&-&-&-1.70\\
\multicolumn{8}{l}{~~For Condition (CBT)}\\
~~~~Intercept 3&-&0.42&-&&-&0.84&-\\
    ~~~~Alg. Mech.&-&-&-0.32&&-&-&0.56\\
    ~~~~Calc. Mech.&-&-&-0.80&&-&-&-0.27\\
    ~~~~Calc. E \& M&-&-&3.27&&-&-&2.40\\\noalign{\smallskip}
\multicolumn{8}{c}{Level 1 and Level 2 Variance}\\ \hline \noalign{\smallskip}
Intercept 1&135.46&135.22&135.61&&95.12&93.56&93.53\\
Level - 1&125.66&125.05&124.65&&98.58&98.08&97.24\\\noalign{\smallskip}
\multicolumn{8}{c}{Level 3 Variance}\\ \hline \noalign{\smallskip}
Int.1/Int.2&4.16&4.35&0.74&&1.01&1.4&0.6\\
Int.1/Cond. (CBT)&-&0.89&0.89&&-&4.81&3.45\\
Post/Int.2&5.51&5.25&2.43&&0.81&0.44&0.23\\
Post/Cond. (CBT)&-&1.94&1.92&&-&2.63&3.12\\ \hline\noalign{\smallskip}
Total Level 3&9.67&12.43&5.98&&1.82&9.28&7.4\\\noalign{\smallskip}
Total Variance&270.79&272.7&266.24&&195.52&200.92&198.17\\
\noalign{\smallskip}\hline 
\multicolumn{8}{l}{*** p\textless0.001. ** p\textless0.01. * p\textless0.05. }
\end{tabular}
}
\end{table}

\begin{figure}
\includegraphics[width=1\columnwidth]{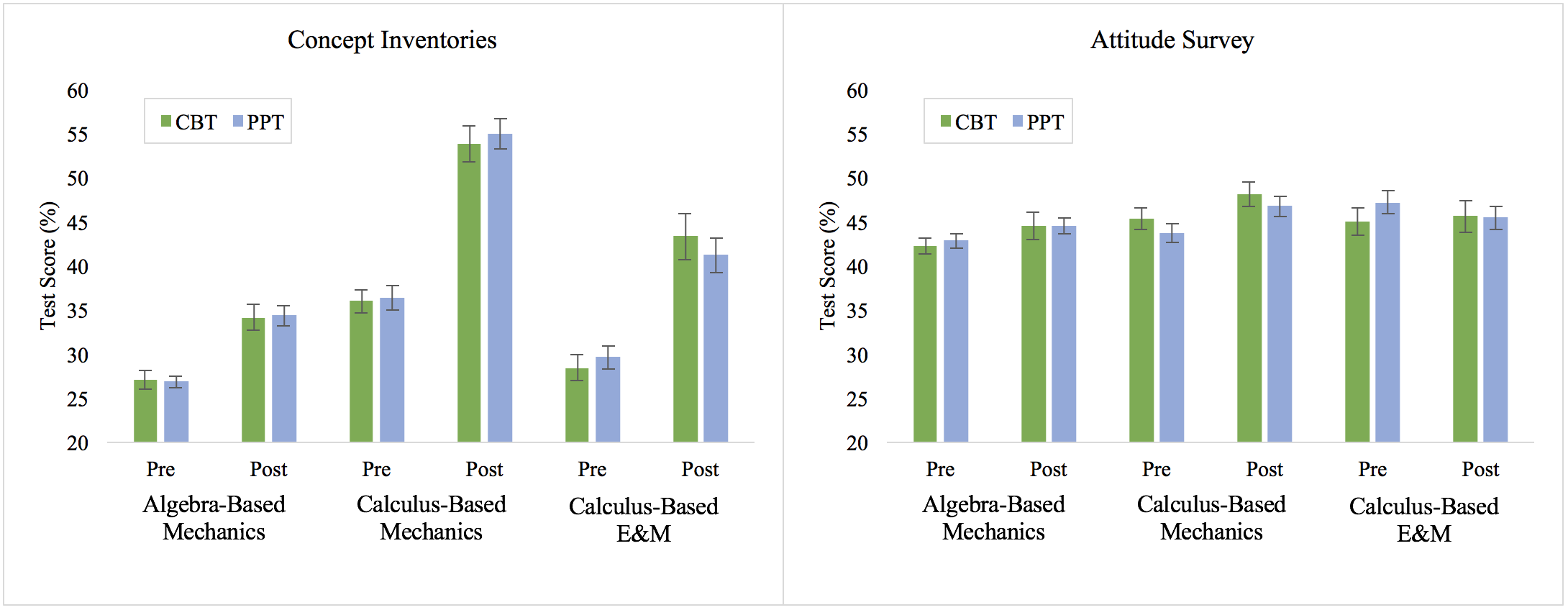}
\caption{A comparison between CBT and PPT administered concept inventories and attitudinal surveys based on  AS Model 3 and CI Model 3. Error bars are 1 standard error. All of the differences between CBTs and PPTs were small, none of the differences were statistically significant, and neither assessment condition was consistently higher than the other. These results indicate that there is no difference in performance between assessments administered as CBTs and those administered as PPTs.
}
\end{figure}

\section{Discussion}
Our HLM models indicate that there is no meaningful difference in scores on low-stakes RBAs between students who completed the RBA in class as a PPT and those who completed the RBA online outside of class as a CBT. Differentiating between CBT and PPT in the models increased the variance in the models, indicating that assessment condition (CBT vs PPT) is not a useful predictor of student scores. The differences between the models' predicted scores for students on both the CI and AS for the PPT and CBT conditions were very small, did not consistently favor one assessment condition over the other, and were not statistically significant. These similarities indicate that instructors and researchers can use online platforms to collect valuable and normalizable information about the impacts of their courses without concerns about the legitimacy of comparing that data to prior research that was collected with paper-and-pencil tests.
\par	In terms of participation, we found that our participation data were comparable to prior research using physics RBAs across several dimensions, including genders and grades. We found that when faculty do little to motivate students to complete online low-stakes assessments, students are much less likely to participate than they are on in-class assessments. Our models show that if faculty follow all of our recommended practices, reminding students in class and online to participate and offering credit for participation, student participation rates for CBT posttests match those for PPT posttests. We focus on the posttests rather than the pretests because the participation rates are lowest on the posttest and they contain important information about the effects of the course.  Our findings align with prior research into student participation on other online surveys, such as end-of-course evaluations. These findings indicate that, with intention, faculty can save class time by transferring their low-stakes RBA administrations from in-class PPTs to out-of-class CBTs without lowering their student participation rates. 
\par	The meaningful differences in participation rates across both student course grades and gender in this study are consistent with what we found reported in prior studies. These differences in participation rates indicate that the missing data in this study, and likely in any study using low-stakes assessments, are not missing at random. We expect that our use of HMI minimized the bias that this introduced into our performance analysis. However, we are not aware of any studies that have explicitly looked at how missing data affect results in studies using low-stakes assessments. Given the frequency with which RBAs are used to assess the effectiveness of college STEM courses, the skew that missing data introduce warrants further investigation.

\section{Conclusion}
Online out-of-class administration of RBAs can provide participation rates and performance results equivalent to in-class paper-and-pencil tests. Instructors should reduce the logistical demands of administering RBAs by using online platforms, such as the LASSO platform, to administer and analyze their low-stakes assessments. paper-and-pencil tests take up already-limited class time and require instructors to use their own time to collect, score, and analyze the assessments. All of these tasks can be easily completed by online platforms, leaving instructors with more time to focus on using the results of the assessments to inform their instruction. Simplifying the process of collecting and analyzing RBA data may lead more instructors to gather this information. By facilitating instructors' examination of their students' outcomes, online platforms may also lead more instructors to start using research-based teaching practices that have been shown to improve student outcomes.
\par	Large-scale data collection with online platforms can also provide instructors with several additional benefits. The platforms can integrate recommended statistical practices, such as Multiple Imputation to address missing data, that most individual instructors do not have the time or expertise to implement. The large scale of the data collection can also be used to put instructors' student outcomes in the context of outcomes in courses similar to their own. Furthermore, analysis could identify teaching practices that the instructor is using that are making their course above average, or practices that they could adopt to improve their outcomes. For example, \url{https://www.physport.org/} is a website that assists faculty in analyzing their existing physics RBA data. The website has a Data Explorer tool that provides instructors with an evaluation of their assessment results and has a series of articles describing highly effective research-based teaching practices that instructors can use to improve student outcomes. 
\par	In addition to supporting instructors, large-scale data collection using online platforms has significant advantages for researchers. It allows investigations into how the implementation and effectiveness of pedagogical practices vary across institutions and populations of students. Large sample sizes provide the statistical power required to investigate differences between populations of students (e.g. gender or ethnicity/race) that would not be possible in most individual courses due to small sample sizes. Online platforms also allow researchers to disseminate new assessments that they are developing so that those instruments can be evaluated across a broad sample of students. Many existing instruments were developed in courses for STEM majors at research-intensive universities with STEM PhD programs, and it is unclear how effective these instruments are for assessing student outcomes at other types of institutions and in courses for non-STEM majors. Online platforms can facilitate analysis of the validity of existing RBA across broad samples of students from all institution types. 
\par	Online data collection and analysis platforms, such as LASSO, are relatively new and have the potential to alter instructor and researcher practices. While it is not known how the transition from PPT to CBT will impact all RBAs, our findings provide strong evidence that two of the most common concerns with digitizing low-stakes RBAs -- shifts in student participation and performance -- are not borne out by the data. Based on the results of our analyses, we recommend that instructors consider using free online RBA administration platforms in conjunction with our four recommended practices for CBTs.

\section{Limitations}
This study only examines courses in which students completed a single low-stakes RBA online at the beginning and end of the course. Excessive measurement would likely decrease student participation, performance, and data quality. Higher-stakes assessments would likely incentivize the use of additional materials (e.g. the internet, textbooks, or peers) not available for tests administered in class. It is also possible that the institution where the study was conducted and the populations involved in the study are not representative of physics students or courses broadly. However, the study included three different courses encompassing both calculus-based and algebra-based physics sequences, which supports the generalizability of results to many populations of students. 
\par	Comparisons of CBT and PPT administered assessments may also be impacted by missing data. Our use of Hierarchical Multiple Imputations (HMI) mitigates the impacts of missing data, but studies that use listwise deletion to address missing data may have different results. The skewing of participation rates by student course grade demonstrates that the data are not missing completely at random and that missing data are therefore non-ignorable.
\section{Directions for Future Research}
The presence and impact of missing data has received little attention in the RBA literature. Most of the studies we reviewed did not provide sufficient descriptive statistics to determine how much data was missing. The majority of studies we reviewed also used listwise deletion to remove missing data and create a matched dataset. Statisticians have long pointed out that the use of listwise deletion is a poor approach to addressing missing data. Our results and the prior studies we examined that provided sufficient information to assess  student participation all indicate that male students and students with lower course grades are less likely to participate in research-based assessments. This skewing of data is likely being amplified through the use of listwise deletion and could have significant impacts research findings. If only the highest performing students reliably participate in an assessment, then the analysis of course data will only indicate the impact on high-performing students and will not be representative of the entire class. We expect that our use of HMI with assessment scores and course grades mitigates the impact on our analysis of the skew in the data. However, almost all studies in Discipline Based Education Research use matched data and do not use appropriate statistical methods for addressing missing data. Future work to measure the impact of missing data and associated data analysis techniques is needed to bring attention to the impact of these issues and provide guidance on methods for limiting their effects.
\par Many institutions are moving to online data collection for their end-of-course evaluations because this streamlines the collection and analysis of student responses. However, instructors are finding that students are much less likely to participate in these surveys than in traditional in-class paper-and-pencil surveys \citep{Dommeyer2004, Stowell2012, Nulty2008, Nair2008, Goos2017}. These surveys often act as the primary methods for institutions to evaluate the effectiveness of instructors and therefore play an important role in retention and promotion decisions. Our results indicate that providing multiple reminders to complete the surveys and participation credit for completing the surveys can dramatically increase participation rates on course evaluations administered online outside of class.

\nocite{Author,Kost2009a,Kost-Smith2010a,Cahill2014}

\bibliography{RIHE.bib}
\end{document}